# A syllogistic system for propositions with intermediate quantifiers

Pasquale Iero[1], Allan Third[2], and Paul Piwek[1]

[1]School of Computing and Communications, The Open University, UK
[2]Knowledge Media Institute, The Open University, UK
*{pasquale.iero,allan.third,paul.piwek}@open.ac.uk*

**Abstract**

This paper describes a formalism that subsumes Peterson's intermediate quantifier syllogistic system, and extends the ideas by van Eijck on Aristotle's logic. Syllogisms are expressed in a concise form making use of and extending the Monotonicity Calculus. Contradictory and contrary relationships are added so that deduction can derive propositions expressing a form of negation.

## 1 Introduction

### 1.1 Aim/Purpose

Monotonicity Calculus is one possible component of Natural Logic (Moss 2012). Monotonicity Calculus can be used to formalise Syllogistic Logic, as shown by Van Eijck (2005).

Syllogistic Logic is a logic based on quantified propositions expressed on terms, given two quantified propositions, called premises. For instance from "Some Athenians are women" and "All women are humans", a third one can be inferred with certainty "Some Athenians are humans".

Monotonicity Calculus uses the order induced by set containment and the monotonicity behaviour of quantified sentences to express logical implication. For instance, the quantifier "All" is downward monotonic in its left-hand argument and upward in its right hand argument, see e.g. (Van Benthem 2007), exemplifying the patterns

$$\downarrow \text{MON } QAB\ A' \subseteq A\ /\ QA'B \qquad (1)$$

That is, if "All humans are mortals", "All women are mortals", and

$$\text{MON} \uparrow QAB\ B \subseteq B'\ /\ QAB' \qquad (2)$$

That is, if "All humans are mortals", "All humans are living beings".

In this paper we extend the Monotonicity Calculus, so that it captures not only the traditional Aristotelian syllogistic inferences on propositions with 'all', 'no', 'some' and 'some ... not', but also inferences involving additional quantifiers such as 'almost all', 'few', 'most', 'most ... not', 'many' and 'many ... not'. These are referred to by Peterson (2000) as intermediate quantifiers. This naming reflects their position in the 'extended square of opposition' (see Fig. 1). The logic system we propose is based on an ordering of quantifiers and operators that capture the relations present in the graph structure of the extended square of opposition. That is we:

1. consider negation relationships, i.e. proposition negation and contrary and contradictory inference between quantified propositions, and intermediate quantifiers.

2. provide a generalisation of syllogistic logic so that we subsume Aristotle's and Peterson's systems, under certain assumptions on non-empty terms.

3. provide a formalism whose components can be modified in order to make the system more expressive (e.g. by using more quantifiers, by using relationships as parameters of quantifiers, etc.).

Only linear ordering between quantified propositions, based on logical implication, is studied.

## 1.2 Importance

The main contribution of this paper is in proposing a logic formalism based on pairs of propositions expressed by intermediate quantifiers and their negation. Inference is obtained by ordering properties of quantified propositions and by two operators that capture the relations of the "extended square of opposition". This square is a graph (Figure 1) that expresses the implication, contradiction and (sub)contrariness relationship between propositions.

Apart from its theoretical interest, the proposed system may have practical applications. For instance, we are currently working on implementing a question generation system (Iero 2017) that can automatically generate comprehension questions from text that includes intermediate quantifiers. An inference engine based on the logical system proposed in this paper is part of the question generator that we are developing.

## 2 Related Work

Natural Logic dates back to Aristotle's 4th century BC syllogisms (Karttunen 2015). Charles Sander Peirce studied a linguistic phenomenon linked to this specific logic, called monotonicity, in the context of formal languages. Natural Logic was revived by the studies of Johan van Benthem and his student Victor Sánchez-Valencia in the 1990s (Valencia 1991). A Natural Logic approach is based on the idea that deciding whether inference is logically sound can be done using syntactic features, as opposed to formal logic approaches where a logical form is derived and reasoning is performed based on it. Although van Benthem crucially linked it with categorial grammars, MacCartney and Manning (2007) show that other formalisations can be used instead.

Parsons (2017) gives a historical context to the traditional square. The square of opposition connects quantified propositions by contrary relationship, when two propositions cannot be true together, subcontrary relationship, when two propositions cannot both be false together and contra- dictory relationship, when two propositions have opposite truth values. Van Eijck (2005) gives a system expressed in terms of monotonicity, symmetry and existential import, that is the assumption that the terms exist, for Aristotle's syllogistic logic. In this way syllogisms are connected to modern studies on Natural Logic, beginning a path of unification between ancient logic and modern theories of logic in language. However, in van Eijck's paper, the inferences that can be obtained by contradictory and contrary relationships are not taken into account (as for instance deriving "It is not the case that no woman is mortal" from "Some woman is mortal").

## 3 Description Of the System

Our system is based on Peterson's intermediate quantifiers theory (Peterson 2000).

Making use of the ordering properties of quantifiers, we propose a way to extend the Monotonicity Calculus with Peterson's intermediate quantifiers. This requires two new operators: one for mirroring the orderings and one for relating elements from one ordering to the other (i.e. from affirmative to negative or vice versa).

### 3.1 The Set of Symbols

The class of symbols describing the syllogistic language can be subdivided in four groups.

1. Terms
2. Quantifiers
3. Negation
4. Syntactic Delimiters

#### 3.1.1 Terms

A sequence of alphanumeric characters in the Roman alphabet starting with a capital letter denotes a term, that is a subject or a predicate. Also an underscore character can be used in the sequence denoting a term (e.g. "Document_writer"). For instance "Man", "Mortal" are terms. "men" and "mortal" are not terms.

### 3.1.2 Quantifiers

Any sequence of alphanumeric roman characters in lowercase starting with letter and possibly containing an underscore ("_") denotes a quantifier. The behaviour of quantifiers is defined separately.

Although, in general, any number of quantifiers can be defined in the language, only a specific number of them will be given for this version of the logic. This version of the logic will contain: "*all*", "*no*", "*some*", "*some_not*", "*almost_all*", "*most*", "*many*", "*few*", "*most_not*", "*many_not*".

### 3.1.3 Syntactic Delimiters

Brackets are used to separate terms from quantifiers. They do not have any linguistic nor logical value, they are just meant to improve human readability of the logic.

### 3.1.4 Negation

Proposition negation is expressed by the symbol '∼' in front of a proposition and it is read as "It is not the case that". For instance the proposition

$$\sim all(\text{Men})(\text{Astronauts}) \quad (3)$$

should be read as "It is not the case that all men are astronauts", that can be simplified in "Not all men are astronauts".

## 3.2 The Grammar

In this section we give a categorial grammar (Lambek 1958) for the fragments of the syllogistic language that we will be using. Forward application only is sufficient to prove that a sentence is well formed.

We define the set of quantifiers

$$q : (Pp/Pr)/Pr \text{ where } q \in \{all, \\ no, \\ some, \\ some\_not, \\ almost\_all, \\ most, \\ many, \\ few, \\ most\_not, \\ many\_not\} \quad (4)$$

the vocabulary of terms (nouns, predicates):

$$T : Pr \text{ where } T \in \{\text{Woman}, \\ \text{Man}, \\ \text{Mortal}, \dots\} \quad (5)$$

The proof of the assignment of the type Pp (Prop) to $q(X)(Y)$ follows.

$$\cfrac{\cfrac{\cfrac{q}{(Pp/Pr)/Pr} \quad \cfrac{(X)}{Pr}}{Pp/Pr} \quad \cfrac{(Y)}{Pr}}{Pp} \text{ ( A well formed Proposition )}$$

$q$ is a quantifier, (X), (Y) are terms.

## 3.3 The Definitions

### 3.3.1 Ordering on Quantifiers

**Definition 3.1.** *We define the following ordered set as the set of affirmative quantifiers:*

$$\mathcal{A} = (A : every, \\ P : almost\_all, \\ T : most, \\ K : many, \\ I : some) \quad (6)$$

The ordering is total ($<$) according to the sequence expressed in Equation (6). However it is related to the boolean implication represented in Figure 1. Therefore the sign $\sqsubset$ will be used instead.

**Definition 3.2.** *In a similar way we define the following totally ordered set:*

$$\begin{aligned}
\mathcal{E} = (&E : no, \\
&B : few, \\
&D : most\_not, \\
&G : many\_not, \\
&O : some\_not)
\end{aligned} \quad (7)$$

*as the set of negative quantifiers.*

The ordering of quantifiers is used to express logical implication between the related quantified propositions. That is:

$$q_1 \sqsubset q_2 \models (q_1(a)(b) \to q_2(a)(b)) \quad (8)$$

### 3.3.2 Contraries and Subcontraries

This system is based on the relationships obtained between propositions and their negation. This pair is made of a proposition and its contrary or subcontrary. We connect contrary and sub-contrary pairs with the operator $\widehat{(X)}$. A contrariness operator gives the contrary or subcontrary of a quantifier.

A contrariness operator has the following properties:

$$\widehat{\widehat{A}} = A \quad (9)$$

Also it associates affirmative and negative quantifiers that are of the same cardinality in their respective orderings:

$$\widehat{A} = E, \widehat{P} = B, \widehat{T} = D, \ldots \quad (10)$$

### 3.3.3 Mirroring Operator

The mirroring operator gives the specular element with respect to the centre element of the set. [1] In $\mathcal{A}$ the centre element is $T$, the mirror of the element $A$, denoted by $A^*$, is $I$. Please note

$$((A)^*)^* = A = (I)^* \quad (11)$$

The same operator applies to the set $\mathcal{E}$.

### 3.3.4 Relationship with Inner and Outer Negation

With the mirroring and contrariness operators we can express inner and outer negation. Using the notation specified in Peters and Westerståhl (2006) for inner and outer quantifiers we have:

$$\hat{Q} = (Q\neg) \quad (12)$$

that is the contrary or sub-contrary negation, and

$$(Q)^* = (\neg Q\neg) \quad (13)$$

Please note:

$$\neg Q = (\widehat{Q})^* = (\hat{Q}^*) \quad (14)$$

That is the contradictory negation.

### 3.3.5 Existential Import

Peterson's system assumes existential import for the affirmative and the negative. That is every proposition implies the existence of both the subject and the predicate. This assumption let us maintain the logical implication between quantified propositions and gives the ordering between quantifiers. That is:

$$q_i < q_j \equiv q_i(X)(Y) \to q_j(X)(Y) \quad (15)$$

## 3.4 The inference rules

Syllogisms are subdivided according to four figures that represent the four possible ways the two quantified propositions can match through a term. In what follows we will be using the lowercase Greek letters '$\alpha$','$\beta$' and '$\gamma$' to represent term variables. A detailed derivation of syllogisms according to the following rules will be added to Iero (2018)

This section also contains examples of syllogistic inference for intermediate quantifiers. The examples are not strict in their syntax, sometimes for readability we match a verb with a term containing an implicit subject (e.g. "those who") sometimes we use the verb negation ("do not") in place of the predicate "are not", sometimes we match the singular with the plural (e.g. "vegetable" with "vegetables"). However, nothing hinges on such minor paraphrases.

---

[1] This is equivalent to the complement value in a set defined on an interval between 0 and 1.

### 3.4.1 The Extended Square

Figure 1 displays two kinds of information between quantified propositions that contain the *same* terms in the *same* positions:

- The relationship between truth values of the quantified proposition.
- The order relationship between quantified expressions.

The order relationship is based on the implication of quantifiers. In this sense the ordering can be reduced to the boolean implication.

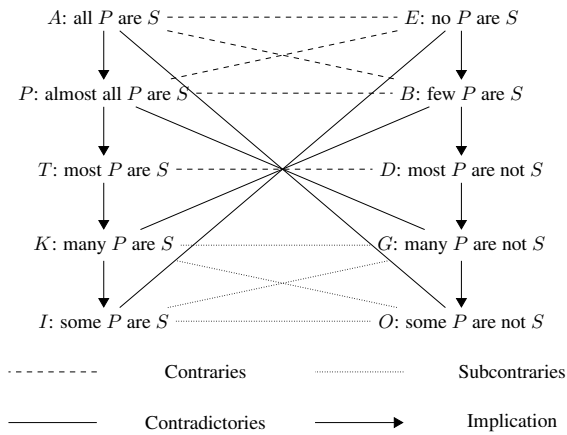

Figure 1: Extended Square of opposition (Peterson 2000)

### 3.4.2 First Figure

The first figure of syllogistic logic is structured as shown below

| Premises | first | $q_{p_1}$ |
| | | $\beta$ |
| | | $\alpha$ |
| | second | $q_{p_2}$ |
| | | $\gamma$ |
| | | $\beta$ |
| Conclusion | | $q_c$ |
| | | $\gamma$ |
| | | $\alpha$ |

We will be expressing what van Eijck calls triggers, $all(\beta)(\alpha)$ and $no(\beta)(\alpha)$, as $\beta \sqsubseteq \alpha$ and $\beta \sqsubseteq \neg\alpha$ respectively. The sign $\sqsubseteq$ designates logical implication. For the set of quantifiers $q \in \mathcal{A}$

$$\frac{q(\gamma)(\beta) \quad \beta \sqsubseteq \alpha}{q(\gamma)(\alpha)} \text{ (I.A)}$$

And indicating $\widehat{q}(\alpha)(\beta)$ as the contrary quantifier of $q(\alpha)(\beta)$

$$\frac{q(\gamma)(\beta) \quad \beta \sqsubseteq \neg\alpha}{\widehat{q}(\gamma)(\alpha)} \text{ (I.E)}$$

**Examples** For reasons of space we will be giving examples only for the first figure. It should be simple enough to apply the schema and draw the equivalent images for the other figures.

As an example of an inference obtained by applying the first figure, we take the sentences: "Most humans can write their names" and "All those who can write their name are able to write". If we match $\alpha$ with "able to write", $\beta$ with "(those who) can write their name" and $\gamma$ with "humans" we can use the first figure (Fig. 2) to derive "Most humans are able to write" given that the quantifier "Most" is $T \in \mathcal{A}$. This inference is illustrated in Fig. 3

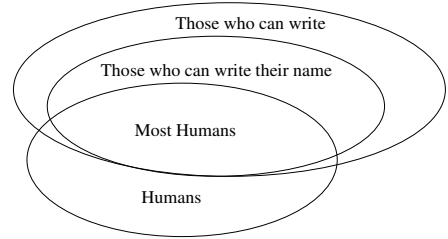

Figure 3: Example of an affirmative syllogism of the first figure

Similarly from "Almost all adults go to work" and "No one of those who go to work sleep until late", by matching $\alpha$ with "(those who) sleep until late", $\beta$ with "(those who) go to work" and $\gamma$ with "adults", (Fig. 4), from $almost\_all = P$ we obtain $\hat{P} = B$ and $B = few$ "Few adults sleep until late" as illustrated in Figure 5. Please note, because of the existential import, we also derive $D, G, O$, that is "Most adults do not sleep until late", "Many adults do not sleep until late" and "Some adults do not sleep until late".

| Premises | first | $Q$ : *All* |
| | | $\beta$: those who can write their name |
| | | $\alpha$: able to write |
| | second | $Q$ : *Most* |
| | | $\gamma$: humans |
| | | $\beta$: (those who) can write their name |
| Conclusion | | $Q$ : *Most* |
| | | $\gamma$: humans |
| | | $\alpha$: are able to write |

Figure 2: Schema of an affirmative syllogism of the first figure

| Premises | first | $q_{p_1}$ : *No one of* |
| | | $\beta$: those who go to work |
| | | $\alpha$: sleep until late |
| | second | $q_{p_2}$ : *Almost all* |
| | | $\gamma$: adults |
| | | $\beta$: (those who) go to work |
| Conclusion | | $q_c$ : *Few* |
| | | $\gamma$: adults |
| | | $\alpha$: sleep until late |

Figure 4: Schema of a negative syllogism of the first figure

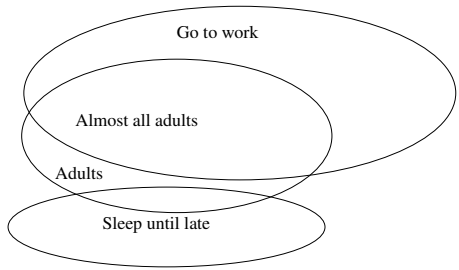

Figure 5: Example of an negative syllogism of the first figure

| Premises | first | $q_{p_1}$ |
| | | $\beta$ |
| | | $\alpha$ |
| | second | $q_{p_2}$ |
| | | $\gamma$ |
| | | $\alpha$ |
| Conclusion | | $q_c$ |
| | | $\gamma$ |
| | | $\beta$ |

Given $q_1 \in \mathcal{E}$ The two inference rules are the following

$$\frac{q_1(\gamma)(\alpha) \quad \beta \sqsubseteq \alpha}{q_1(\gamma)(\beta)} \text{ (II.A)}$$

and given $q_2 \in \mathcal{A}$

$$\frac{q_2(\gamma)(\alpha) \quad \beta \sqsubseteq \neg\alpha}{\widehat{q_2}(\gamma)(\beta)} \text{ (II.E)}$$

### 3.4.3 Second Figure

The second figure of syllogistic logic is structured as shown below

### 3.4.4 Third figure

The third figure of syllogistic logic is structured as shown below

| Premises | first | $q_{p_1}$ |
| | | $\gamma$ |
| | | $\alpha$ |
| | second | $q_{p_2}$ |
| | | $\gamma$ |
| | | $\beta$ |
| Conclusion | | $q_c$ |
| | | $\beta$ |
| | | $\alpha$ |

We have as inference rules, for $q_1, q_2 \in \mathcal{A}$

$$\frac{q_1(\gamma)(\alpha) \quad q_2(\gamma)(\beta) \sqsubseteq ((q_1)(\gamma)(\beta))^*}{I(\beta)(\alpha)} \text{ (III.A)}$$

and for $q_1 \in \mathcal{E}$ and $q_2 \in \mathcal{A}$

$$\frac{q_1(\gamma)(\alpha) \quad q_2(\gamma)(\beta) \sqsubseteq (\widehat{q_1}(\gamma)(\beta))^*}{O(\beta)(\alpha)} \text{ (III.E)}$$

For instance for $q_1 = B$

$$\widehat{B} = P$$

and

$$P^* = K$$

Therefore for

$$q_2 \sqsubseteq K \Rightarrow q_2 \in \{A, P, T, K\}$$

we obtain $O$ from (III.E)

### 3.4.5 Fourth Figure

The fourth figure of syllogistic logic is structured as shown below

| Premises | first | $q_{p_1}$ |
| | | $\alpha$ |
| | | $\beta$ |
| | second | $q_{p_2}$ |
| | | $\beta$ |
| | | $\gamma$ |
| Conclusion | | $q_c$ |
| | | $\gamma$ |
| | | $\alpha$ |

And given $q_1, q_2 \in \mathcal{A}$ we have the following inference rules:

$$\frac{q_1(\alpha)(\beta) \quad \beta \sqsubseteq \gamma}{I(\gamma)(\alpha)} \text{ (IV.A)}$$

$$\frac{\alpha \sqsubseteq \beta \quad \beta \sqsubseteq \neg\gamma}{E(\gamma)(\alpha)} \text{ (IV.Æ)}$$

$$\frac{\alpha \sqsubseteq \neg\beta \quad q_2(\beta)(\gamma)}{O(\gamma)(\alpha)} \text{ (IV.E)}$$

### 3.4.6 Contradictories

The contrariness/sub-contrariness relationship in the extended square in Figure 1 is formalised as follows:

$$\frac{q}{\sim (\hat{q})^*}$$

and

$$\frac{\sim q}{(\hat{q})^*}$$

### 3.4.7 Chain of implication

The extended square gives logical implication for affirmative and also for negative quantifiers (Figure 1). Interpreting vertical arrows as logical implication means that if a proposition is true all those that follow it are true and if a proposition is false, all propositions that precede it are false. That is

$$\frac{q_i \quad q_i \sqsubset q_j}{q_j} \text{ (ExI.A)}$$

and

$$\frac{\sim q_i \quad q_j \sqsubset q_i}{\sim q_j} \text{ (ExI.E)}$$

### 3.4.8 Metaresult

Syllogisms never derive a negated proposition, a proposition starting with "It is not the case that". These propositions are obtained by applying the relationships encoded by the square of opposition. Therefore we can say that any quantified proposition that is before the contradictory of the result in the linear sequence of quantifiers, both $\mathcal{A}$ and $\mathcal{E}$, must be false, or be $q_i(X)(Y)$ the inferred quantified proposition, $q_c(X)(Y)$ its contradictory, and $q_x(X)(Y)$ a quantified proposition that is before the contradictory,

$$\begin{aligned}q_x(X)(Y) \sqsubseteq q_c(X)(Y) \vdash \\ \text{it follows that not } q_x(X)(Y)\end{aligned} \quad (16)$$

## 3.5 A theorem

This section introduces a theorem that can be obtained immediately from the system described in this chapter.

### 3.5.1 Restriction

The system so far described is composed of 5 quantities and therefore called the 5-quantities system. This system subsumes the classical syllogisms, composed by 2 quantities. However, it is interesting to ask whether the classical syllogism can be derived with a 2-quantities system with the formalism given, that is with

**Definition 3.3.** *The 2-quantities system contains the 2 affirmatives:*

$$\mathcal{A}_2 = \{A, I\} \qquad (17)$$

*and the 2 negatives:*

$$\mathcal{E}_2 = \{E, O\} \qquad (18)$$

**Theorem 1.** *Syllogistic rules and $\mathcal{A}_2, \mathcal{E}_2$ derives classical syllogisms with existential import*

*Proof.* The inference rules do not change, the sets are $\mathcal{A}_2, \mathcal{E}_2$. What changes is the mirroring operator, since the sets are of even cardinality. The operator, in this case is $(A)^* = I$ and $(E)^* = O$. This by enumeration derives Aristotle's syllogisms with existential import using Monotonicity rules, one of the components of Van Eijck's system. However, symmetry is not used. The operators we give obtain an equivalent result. □

## 4 Conclusion and further work

We have described a logical system that amends and extends the Monotonicity Calculus of Van Eijck (2005). This allows us to capture the syllogistic and square of opposition inferences for Peterson's intermediate quantifiers. Rather than enumerate the syllogisms, we present a small number of inference schemes from which instances of the syllogisms are derivable. This deduction system is part of the design of a logic engine of an automated question generation system which design is outlined in Iero (2017). We are planning to extend the system using more intermediate quantifiers, and by adding relationships, in order to increase the expressive power of our logic engine.

## 5 Acknowledgements

I wish to thank my family, that is my mother and my brother, who are letting me do this research sacrificing their personal time.